\documentclass[english,showpacs,prl,twocolumn]{revtex4}
\usepackage[T1]{fontenc}
\usepackage[latin9]{inputenc}
\usepackage{float}
\usepackage{amssymb}
\usepackage{graphicx}

\makeatletter
\@ifundefined{textcolor}{}
{%
 \definecolor{BLACK}{gray}{0}
 \definecolor{WHITE}{gray}{1}
 \definecolor{RED}{rgb}{1,0,0}
 \definecolor{GREEN}{rgb}{0,1,0}
 \definecolor{BLUE}{rgb}{0,0,1}
 \definecolor{CYAN}{cmyk}{1,0,0,0}
 \definecolor{MAGENTA}{cmyk}{0,1,0,0}
 \definecolor{YELLOW}{cmyk}{0,0,1,0}
}

\makeatother

\usepackage{babel}
\begin{document}

\title{Self Organization and Self Avoiding Limit Cycles}

\author{Daniel Hexner and Dov Levine}

\affiliation{Department of Physics, Technion-IIT, 32000 Haifa, Israel}
\begin{abstract}
A simple periodically driven system displaying rich behavior is introduced
and studied. The system self-organizes into a mosaic of static ordered
regions with three possible patterns, which are threaded by one-dimensional
paths on which a small number of mobile particles travel. These trajectories
are self-avoiding and non-intersecting, and their relationship to
self-avoiding random walks is explored. Near $\rho=0.5$ the distribution
of path lengths becomes power-law like up to some cutoff length, suggesting
a possible critical state. 
\end{abstract}


\maketitle
When driven periodically, many-body systems display a wide variety
of behavior, including highly complex spatial and dynamical self-organization.
For example, vibrated beds of sand develop extended geometrical structures\cite{melo_hexagons_1995},
a periodically driven damped Frenkel-Kontorova model organizes to
a marginally stable state\cite{tang_phase_1987}, and a perfect flowing
state develops in a simple traffic model\cite{biham_self-organization_1992}.
In some systems a phase transition occurs between a chaotic phase
and a phase which is periodic and slaved to the driving\cite{corte_random_2008,Coppersmith_Boolean_Rev}.
Another example of complex self-organization was seen in a recent
simulation of a periodically sheared granular packing\cite{Royer}:
the system enters a limit cycle, where each individual grain moves
in its own intricate path, ultimately returning to its starting point
at the end of a cycle.

In this Letter, we introduce a minimal model of periodically driven
particles on a lattice. Despite the simplicity of the model, complex
behavior arises, with a steady state exhibiting two salient features:
(1) The (great) majority of the system self-organizes into regions
of half-filling which are invariant under the dynamics, and (2) The
rest of the particles become entrained, moving in periodic orbits
whose paths appear to be both non-intersecting and self-avoiding.
The distribution of the lengths of these paths is narrow for small
densities, but in the vicinity of $\rho=0.5$ it tends towards a power-law
distribution, suggesting the possibility of a phase transition. 

\begin{figure}[H]
\begin{centering}
\includegraphics[scale=0.3]{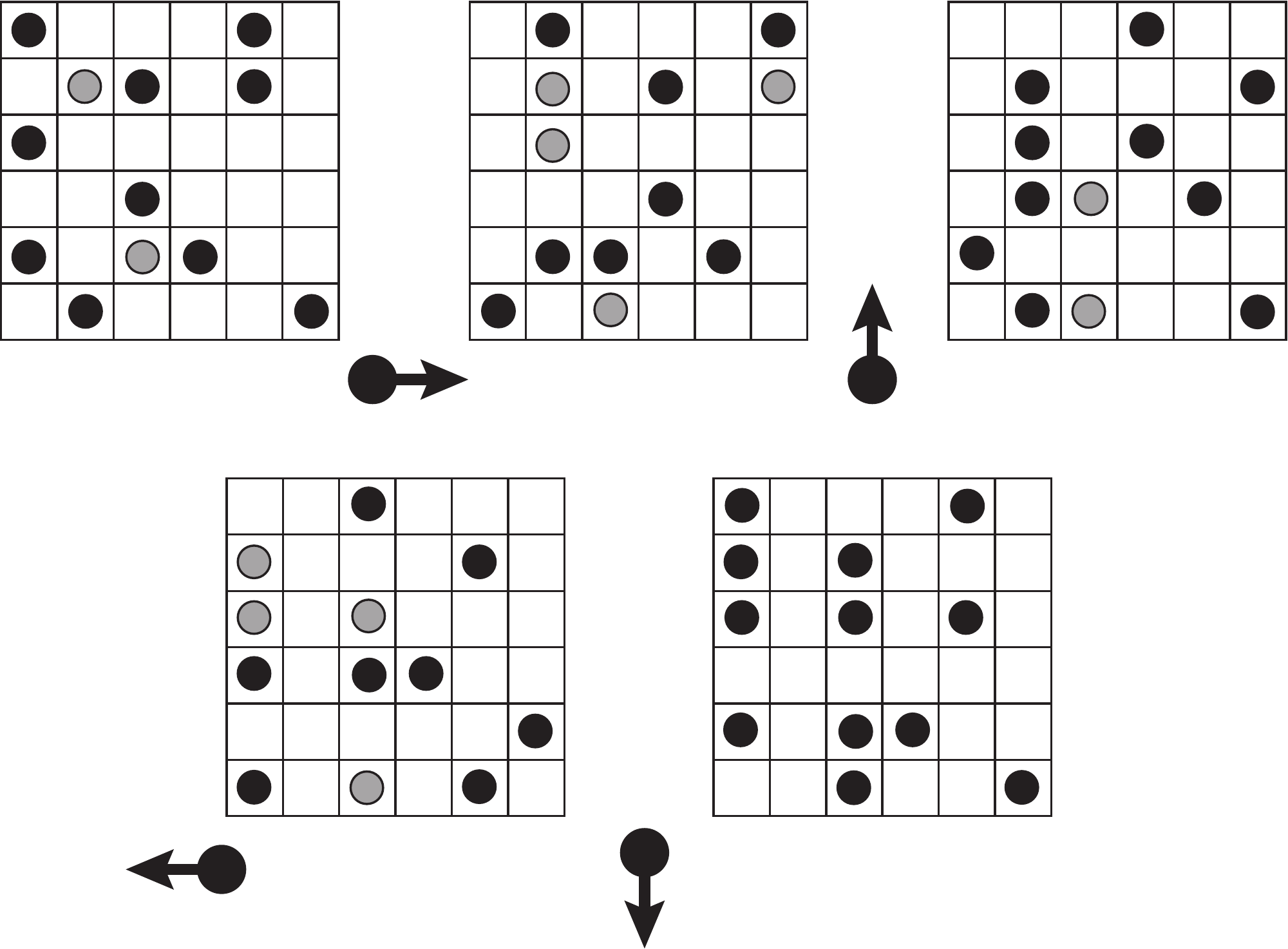}
\par\end{centering}

\caption{An illustration of the one cycle of the model. Black circles mark
unblocked particles that are free to move in the following move, while
grey particles are blocked for the next move. The directions of the
moves are indicated by arrows, and periodic boundary conditions are
employed.\label{fig:the_dynamics-1}}
\end{figure}

In its simplest version, the model is defined on an $L\times L$ square
lattice, where each lattice site may be occupied by up to a single
particle. The boundary conditions are periodic unless specified otherwise.
Each cycle of the dynamics is composed of four moves, in each of which
all the particles which are not blocked translate by one site. The
first move is to the right, the second is up, the third is to the
left, and the last move is down. The updating scheme is parallel:
a particle which is blocked before a move is attempted does not participate
in that specific move. Specifically, in the first move, any particle
which has no neighbor to its immediate right is translated to the
right, while blocked particles do not move; this is then repeated
in the remaining directions, as indicated in Figure \ref{fig:the_dynamics-1}.
If a particle is not blocked at all during a step, its motion is unhindered
and it traces out a square, returning to its original position at
the end of the cycle. 

\begin{figure}
\includegraphics[scale=0.25]{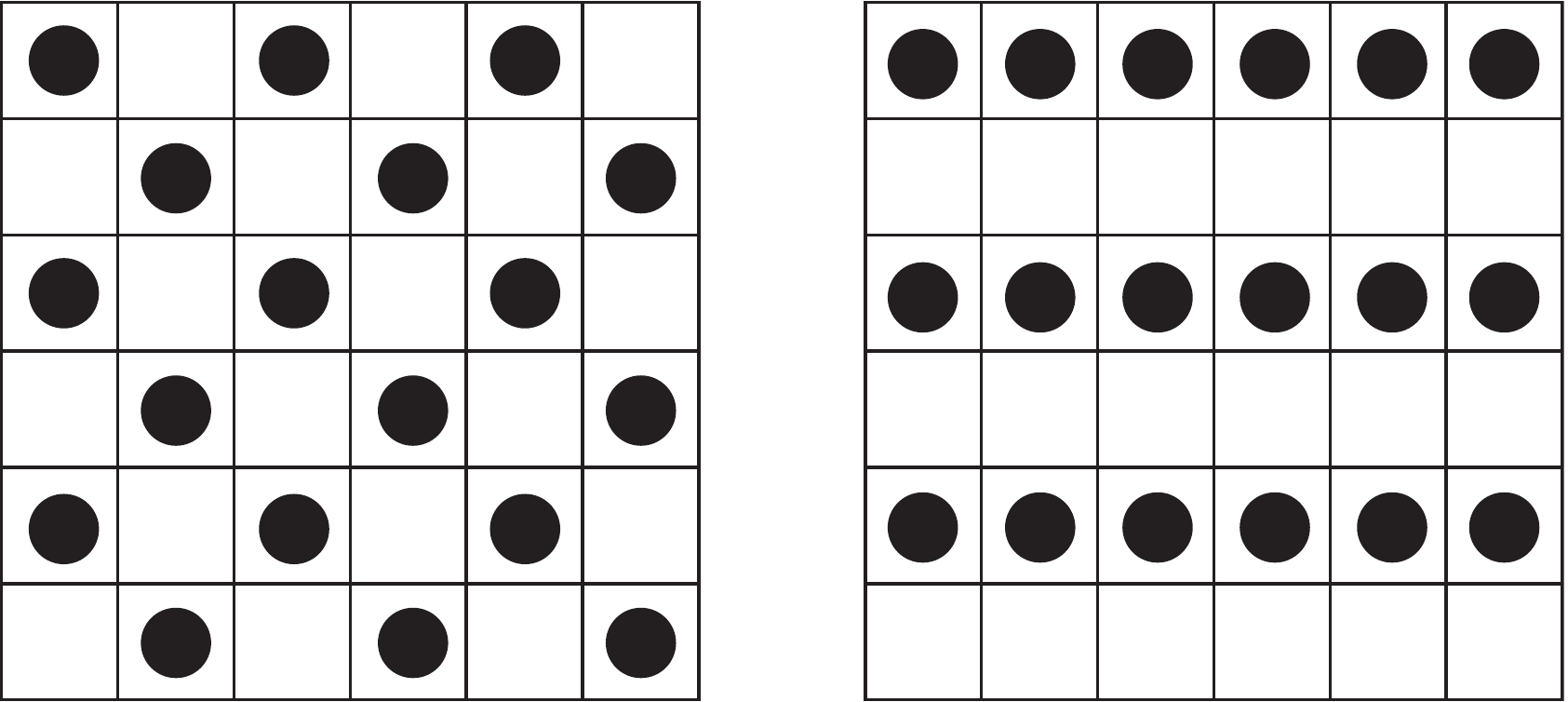}\caption{Two invariant configurations, checkerboard and striped, each at half
filling. Each maps onto itself after one cycle. \label{fig:inv_conf}}
\end{figure}

We typically ``strobe'' the system, comparing configurations before
and after a full cycle. Thus, a particle which was not blocked at
all during a cycle will appear to be stationary, since it returns
to its initial configuration. Note that it is not necessarily the
case that a particle which is blocked for one or more of the moves
will fail to return to its original location. We will refer to particles
or clusters of particles which return to their starting positions
at the end of a cycle as `invariant under the dynamics', or `stationary'.
Figure \ref{fig:inv_conf} shows two such invariant patterns at half
filling.

We have studied the behavior of the model through numerical simulations
starting from random initial conditions for particle densities $\rho\in\left[0,0.5\right]$.
We note that the model is self-dual: because of particle-hole symmetry,
higher densities $\rho\in\left[0.5,1\right]$ are mapped onto the
lower half-interval by $\rho\rightarrow(1-\rho)$. This can be seen
easily at the level of a single move where the motion of a particle
to the right can be envisioned as the motion of a vacancy to the left.
Thus, the behavior at densities $\rho>0.5$ is identical to behavior
at density $1-\rho$ if vacancies are tracked instead of particles.

The steady state behavior of the dynamics is very different from many
other driven systems \cite{corte_random_2008,lubeck_universal_2004,Coppersmith_Boolean_Rev}
in that for the full range of density, the system self-organizes,
displaying no chaotic behavior. At low densities this proceeeds in
a trivial manner with almost all the particles becoming separated,
allowing them to complete a cycle of the motion unhindered by any
blocking, with but a few undergoing periodic motion with a short period.
At these low densities, the rearrangements leading to the steady state
are local in character, as might be expected. At higher densities,
however, the organization to a steady state occurs on long length
scales. This is seen both in the long relaxation dynamics and in the
characteristic formation of large, well-defined regions. 

\begin{figure}
\begin{raggedright}
\includegraphics[clip,scale=0.5]{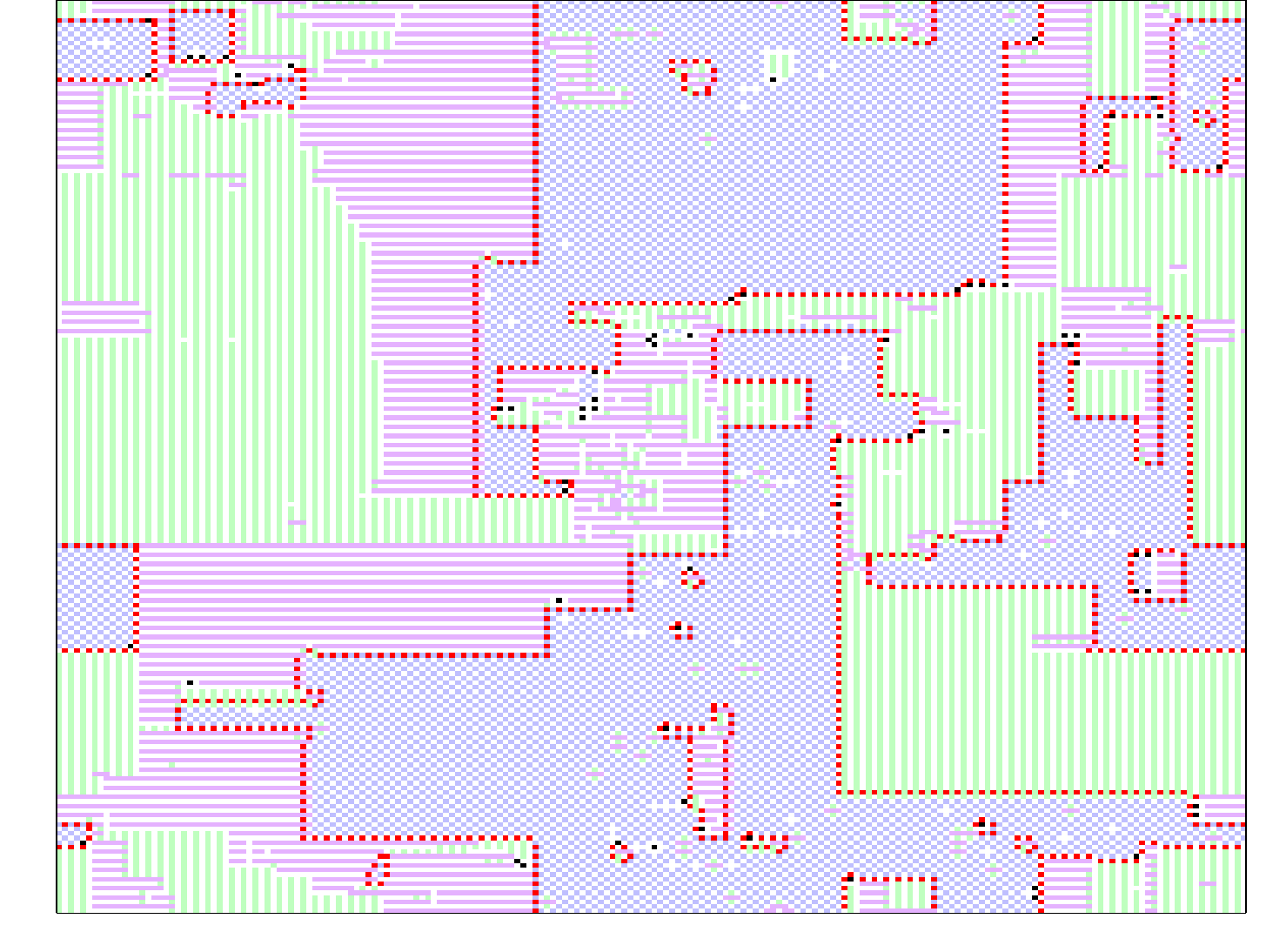}
\par\end{raggedright}

\caption{An example of steady state configuration at $\rho=0.5$. Blue marks
the checkered pattern, green marks the vertical striped pattern and
pink marks the horizontal striped pattern. The paths of mobile particles
are indicated in red. \label{fig:Conff}}
\end{figure}

These regions are invariant under the dynamics, and appear static
under strobing. They always self-organize into three possible patterns:
either vertical stripes, horizontal stripes or a checkerboard pattern
(see Figure \ref{fig:inv_conf})%
\footnote{We also note that each pattern may occur in two possible phases (for
example the vertical stripes can be translated by a single space to
the right), as expected from the particle-hole symmetry.%
}, with occasional point defects. In fact, the whole steady-state configuration
can be viewed as a mosaic of grains of these patterns as shown in
Figure \ref{fig:Conff}. We emphasize that all three of these patterns
have a density of $\rho=0.5$, and that each of them is stable against
the removal or addition of a particle in the bulk (these two operations
are the same due to the particle-hole symmetry). This differs considerably
from directed percolation models which also have invariant configurations\cite{corte_random_2008},
but which are not stable to local perturbations%
\footnote{This may be the reason that our does not display a chaotic phase,
and could be a general requirement for similar models to have no chaotic
phase. %
}.

It is clear from the above that stationary configurations at many
densities exist, being easily obtained by adding (or removing) particles
in the bulk of the different grains. For example, any number of particles
can be added to the checkered pattern at any location and the configuration
will still be stationary. It is therefore surprising that in a broad
range of densities centered around $\rho=0.5$ the system arranges
itself into a mosaic of grains of almost perfectly ordered striped
or checkerboard patterns at half filling. These form as a result of
a dynamical process which we discuss presently.

Not all the particles in the steady-state are stationary; a small
fraction of ``frustrated'' particles move about the system. Remarkably,
these particles are confined to one dimensional directed closed paths
which intersect neither themselves nor other paths, although they
often form very complicated winding curves, especially near $\phi=0.5$;
this is depicted in Figure \ref{fig:Conff}. These particles move
in periodic orbits along their designated paths, with periods which
may be extremely long. 

As seen in Figure \ref{fig:Conff}, there is a close relation between
the invariant grains and the paths of the moving particles, with the
paths appearing on boundaries separating checkerboard and striped
patterns. The motion along each path is unidirectional and occurs
either clockwise or counter-clockwise depending on the pattern in
the interior of the path. If the checkered region is in the interior,
the motion is clockwise, while when the internal region is striped,
the motion is counter-clockwise. The latter case is rare. It is not
the case that motion occurs on every boundary between checkered and
striped phases; in these cases, the boundary is filled with vacancies
($\rho<0.5$) or completely occupied ($\rho>0.5$).

\begin{figure}[H]
\begin{centering}
\includegraphics[scale=0.3]{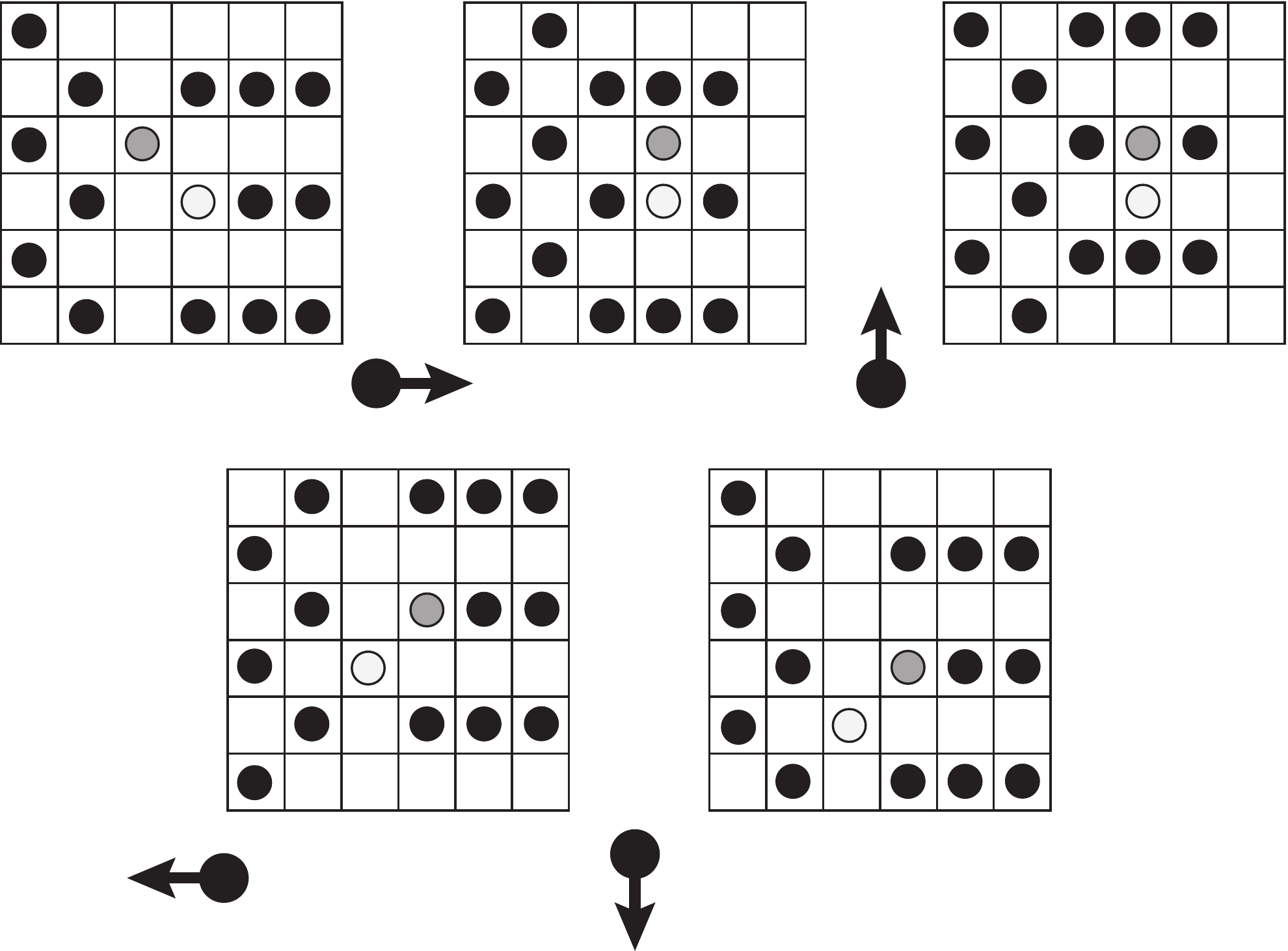}
\par\end{centering}

\caption{Motion of tagged particles at a stripe-checkerboard boundary. Comparing
the state at the end of a cycle with that at the beginning (and regarding
the particles as indistinguishable), gives the illusion of a single
particle being displaced downward by two units. This is, of course,
not possible for a single particle acting alone.\label{fig:bound_dynam}}
\end{figure}

To see how the motion on the stripe-checkerboard grain boundary occurs
we track two particles explicitly in Figure \ref{fig:bound_dynam}.
Note that the motion (for example, vertical jumps of two lattice spacings,
as in the figure) cannot be attributed to a single tagged particle
hopping along the path, but rather through the motion of two particles.
For this reason the paths in figure \ref{fig:Conff} appear as dashed
lines.

The dynamical mechanism responsible for the almost perfect ordering
of the checkerboard and striped patterns is collective in nature,
and occurs in the latter stages of the transient period, when paths
have formed but are not quite stable. In this stage, in addition to
the motion which occurs along paths as in the steady-state, the motion
sometimes penetrates into the grain bulk as a front ``sweeping''
across the grain. In the wake of such a moving front, the nature of
the pattern changes, from striped to checkerboard or \emph{vice versa}
(see supplemental material), and new paths are formed. In this coarsening
process the size of the clusters grows until the steady state is reached
and the paths are stable.

\begin{figure}
\includegraphics[scale=0.6]{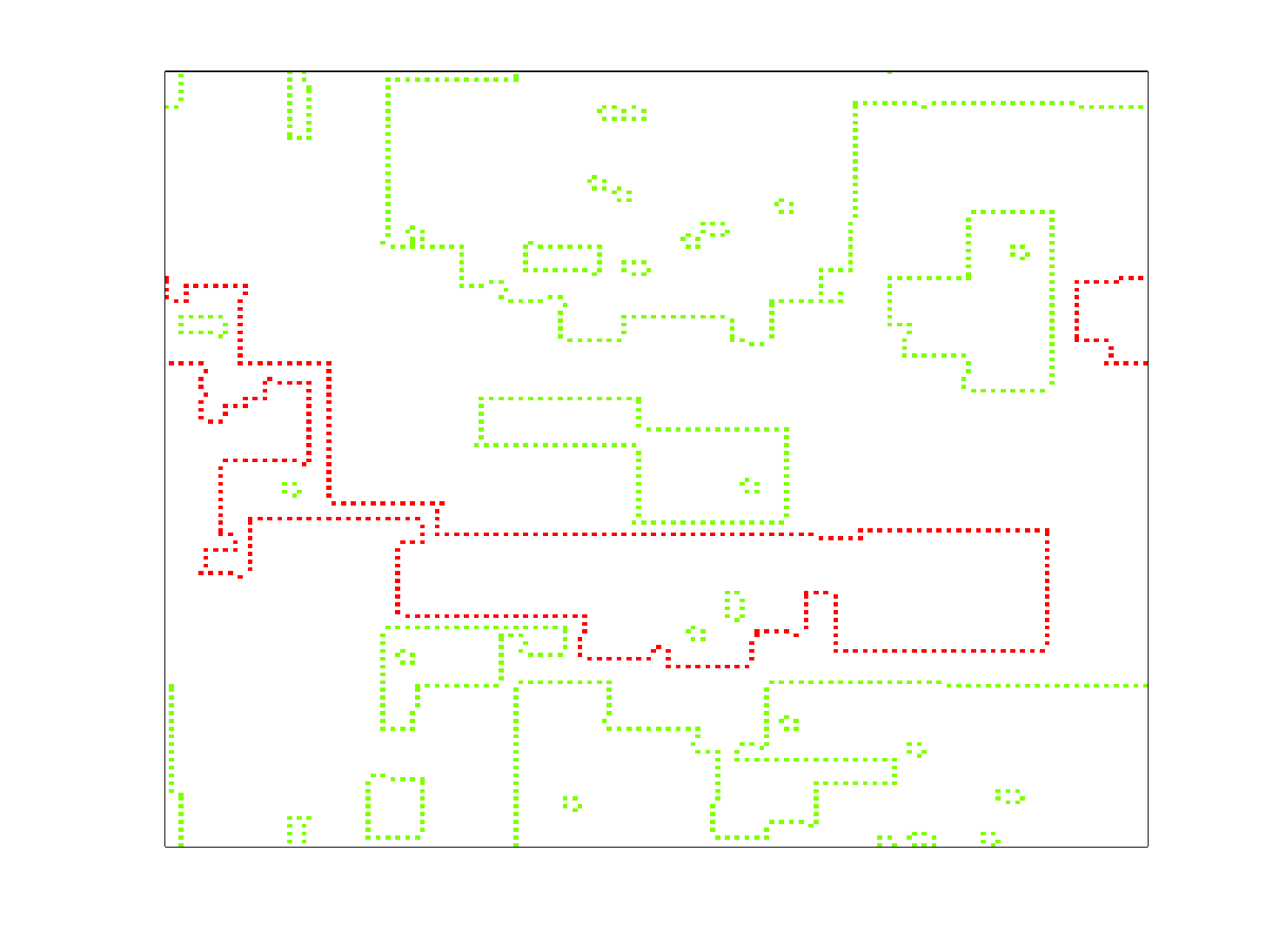}

\caption{An example of the paths traced out by moving particles in the steady
state, with $L=400$ and $\rho=0.5$. The red trajectory marks the
longest path. Note that the dotted nature of the paths is due to the
motion of a pair, as shown in Figure \ref{fig:bound_dynam}. \label{fig:Paths}}
\end{figure}

We begin our quantitative analysis of the steady-state by examining
the paths. These are mapped out numerically by allowing the system
to reach steady-state, and then time-averaging over the moving particles.
In particular, we calculate $\rho_{ij}\equiv\left\langle \left|X_{ij}\left(t+1\right)-X_{ij}\left(t\right)\right|\right\rangle $,
where $X_{ij}\left(t\right)\in\left[0,1\right]$ marks the occupancy
of the site with coordinates $\left(i,j\right)$ at time $t$, with
$t+1$\emph{ }being reckoned at the end of the following cycle. In
the invariant regions, $\rho_{ij}=0$, while along a path, $\rho_{ij}$
is non-zero, and its value is related to the density of particles
along the path. It is worth noting that the value of $\rho_{ij}$$ $
is constant along a path, so that the flux along any given path is
constant. An example the non-zero elements of $\rho_{ij}$ is presented
in Figure \ref{fig:Paths}.

In some instances, two paths touch each other, raising the question
of whether to consider these as independent paths or a single path
with a loop. We argue in favor of the former interpretation for the
following reasons. First, particles traversing touching portions move
in opposite directions, and second, the flux of moving particles is
constant on all points of a given path and typically different from
other paths. There are rare exceptions where two touching paths ``interact'',
with a particle jumping between the two paths. In this case, in the
touching region $\rho_{ij}$ has a value different from that of both
parent paths where there is no touching. This leads us to define a
path as the set of connected points with the same $\rho_{ij}$. An
alternative, not presented here, is to track the location of tagged
particles and measure the paths they follow; this yields the same
qualitative behavior. 

\begin{figure}
\begin{centering}
\includegraphics[scale=0.6]{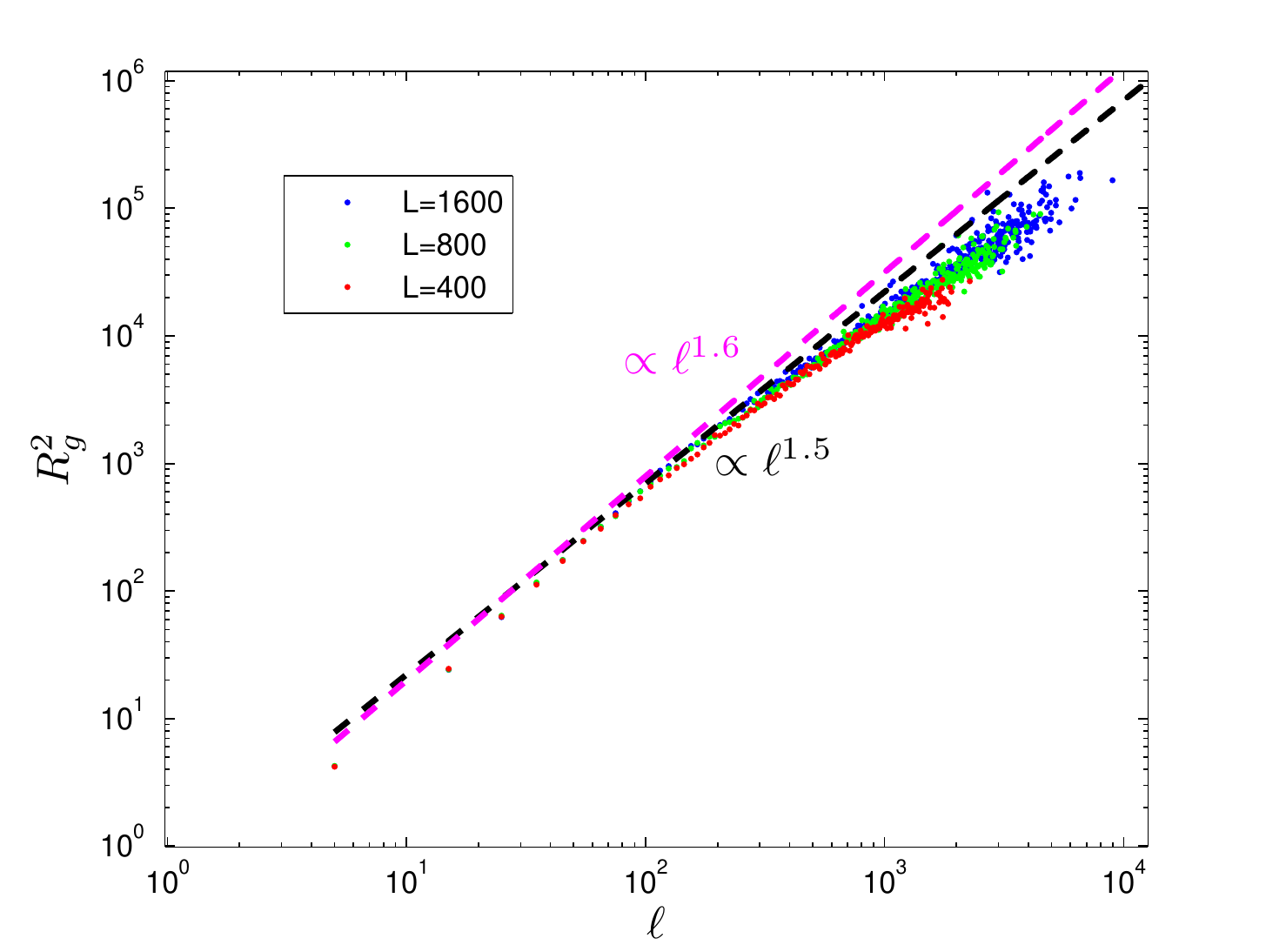}
\par\end{centering}

\caption{Squared radius of gyration as a function of the path length for different
system sizes at $\rho=0.5$. Possible scaling relations $\ell^{2\nu}$
with $\nu=0.75$ and $0.8$ are presented.\label{fig:Radius-of-gyration}}
\end{figure}

As seen in Figure \ref{fig:Paths}, paths are both non-intersecting
and self-avoiding. There are several different ensembles of self-avoiding
paths known, among them simple closed curves, self-avoiding random
walks, and loop erased walks\cite{LoopErased}. These differ in their
fractal dimension, which can be characterized by the scaling of the
radius of gyration $R_{g}=\sqrt{\left\langle \left|\vec{r}-\vec{r}_{0}\right|^{2}\right\rangle }$,
where $\vec{r}_{0}$ is the center of mass of the walk%
\footnote{When measuring the gyration radius, we choose impenetrable walls as
boundary conditions.%
}. The gyration radius scales as a power of the path length $\ell$,
namely, $R_{g}\sim\ell^{\nu}$, where for a regular polygon $\nu=1$,
for a self-avoiding walk $\nu=0.75$\cite{SAW1} and for loop erased
walk $\nu=0.8$\cite{LoopErased}. In Figure \ref{fig:Radius-of-gyration}
we show $R_{g}^{2}\left(\ell\right)$ at $\rho=0.5$. The behavior
of $R_{g}\left(\ell\right)$ is consistent with a power-law with $\nu=0.75$
or $\nu=0.8$, though some deviation occurs for large $\ell$. This
deviation may result from finite size effects such as the boundary
which limits the gyration radius, while another possibility is that
it is due to the interaction between the different paths, causing
crowding. An estimate to the order of magnitude of the linear size
of the cluster at which the crossover occurs can be estimated from
where $R_{g}\left(\ell\right)$ deviates from a power-law; this appears
to be of order of several hundred. 

\begin{figure}
\includegraphics[scale=0.6]{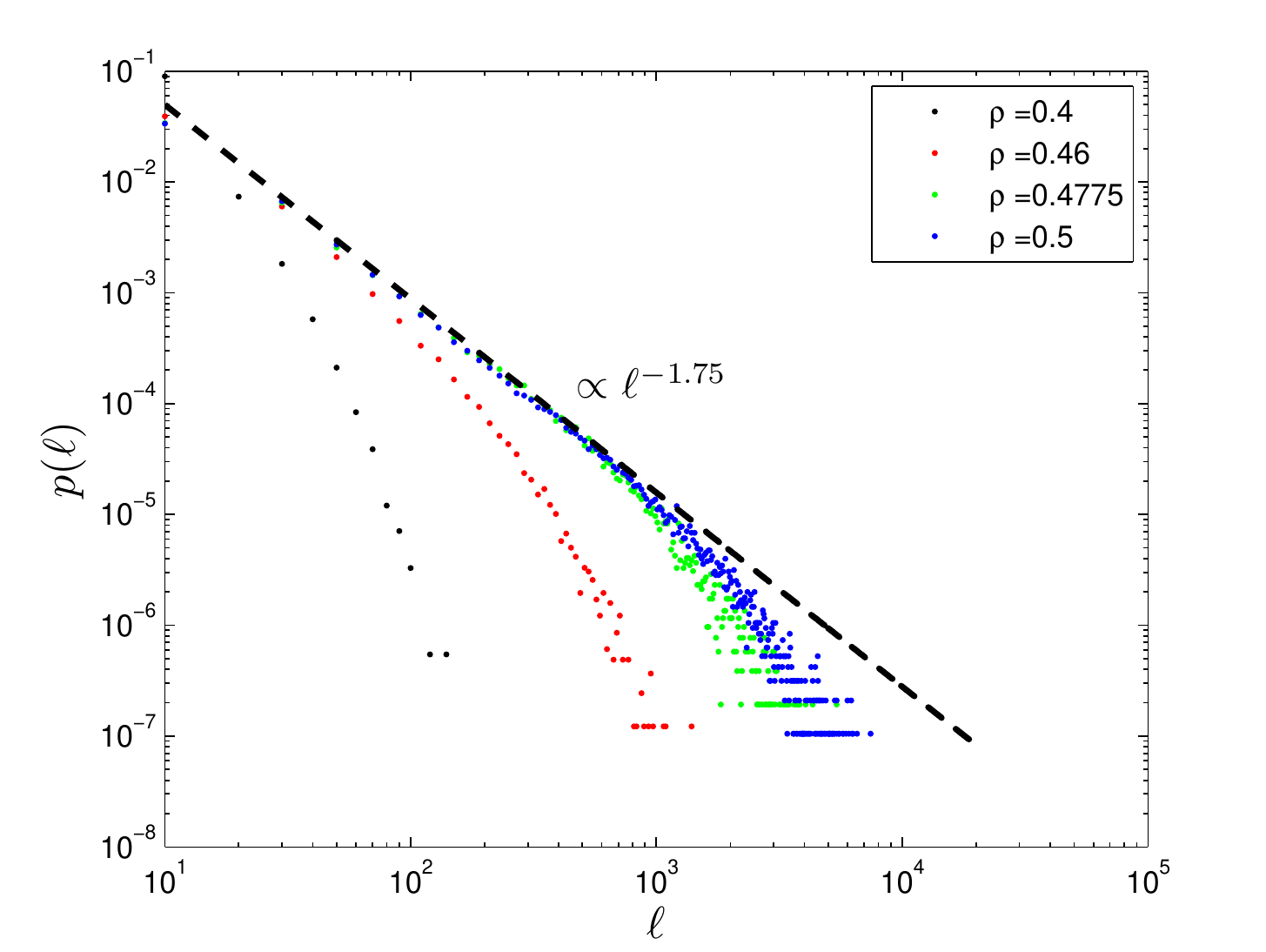}

\caption{The distribution of path lengths for different densities, here $L=1600$.
The black line is $\ell^{-1.75}$. \label{fig:p_of_ell}}
\end{figure}

\begin{figure}
\includegraphics[scale=0.6]{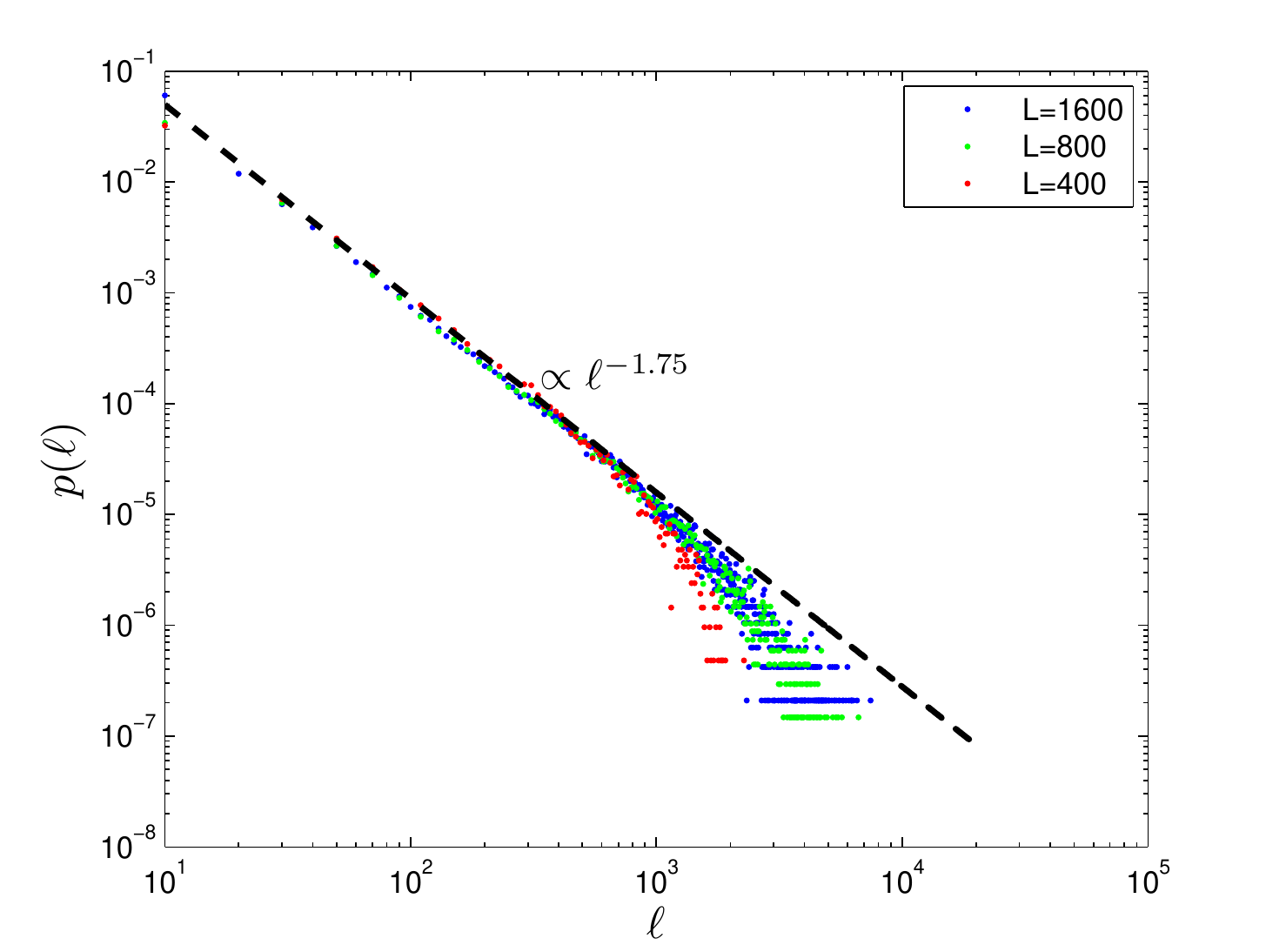}

\caption{The distribution of path lengths at $\rho=0.5$ for different system
sizes. The black line is $\ell^{-1.75}$. \label{fig:p_of_el_0.5}}
\end{figure}

In addition to $R_{g}$, we measured $P\left(\ell\right)$, the distribution
of paths of length $\ell$, for different densities. At low densities,
the fall-off of $P\left(\ell\right)$ at short distances is consistent
with exponential decay. At $\rho\gtrsim0.475$, the decay appears
to be power-law: $P\left(\ell\right)\propto\ell^{-\gamma}$ , where
$\gamma\simeq1.75$, as seen in Figure \ref{fig:p_of_ell}%
\footnote{We note that since paths usually enclose a checkered pattern, the
distribution of checkered grain size can be expected to be a power-law
as well, and we have verified that this is the case.%
}. This power-law behavior persists up to a system size-dependent length,
at which the distribution falls off more quickly. As shown in Figure
\ref{fig:p_of_el_0.5} there is little difference between $L=800$
and $L=1600$ at $\rho=0.5$ with similar behavior at other densities.
Our results are inconclusive as to whether this dropoff is due to
finite size effects or to a finite correlation length, and as such
it is difficult to conclude with confidence whether or not the system
becomes critical in the limit of $L\rightarrow\infty$, or whether
a phase transition occurs or is avoided. 

The value of $\gamma$ is consistent with the probability of occurrence
of a path being inversely proportional to its area: The linear size
of a path scales as $R_{g}$, and if the frequency of occurrence is
inversely proportional to area, the probability density goes as $P\left(R_{g}\right)\propto R_{g}^{-2}$.
Since $R_{g}\propto\ell^{\nu}$, it follows that $P\left(\ell\right)\propto\ell^{-\nu-1}$,
so the exponents are related by $\gamma=1+\nu$. For self avoiding
random walks, $\nu=0.75$, giving $\gamma=1.75$, which is consistent
with our measurements. 

\begin{figure}
\includegraphics[scale=0.6]{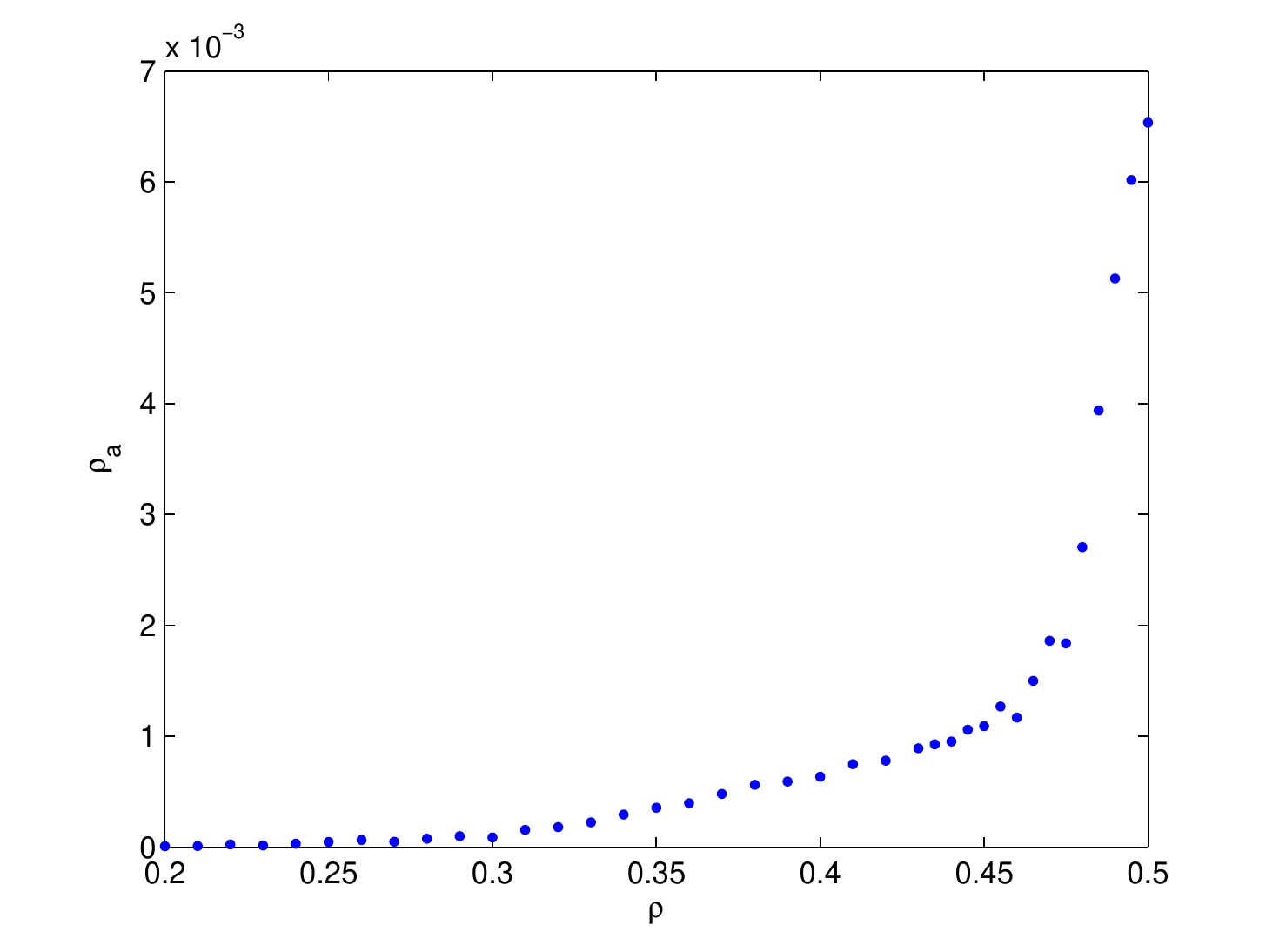}

\caption{The asymptotic density of mobile particles $\rho_{a}\left(t=\infty\right)$
in the steady state for a system of size $L=1000$. \label{fig:RhoA}}
\end{figure}

To gain insight into the dynamics, we measure the density of moving
particles as a function of time, which in analogy to the directed
percolation transitions, we call the active density, denoted $\rho_{a}\left(t\right)$.
For low densities, $\rho_{a}\left(t\right)$ decays to its asymptotic
value $\rho_{a}\left(t=\infty\right)$ in an exponential fashion.
This asymptotic value is shown in Figure \ref{fig:RhoA} as a function
of the density, and shows a steep rise at $\rho\simeq0.475$. As the
density grows, the time scale on which $\rho_{a}\left(t\right)$ decays
grows, perhaps even diverging.

While it is tempting to conclude that there may be a phase transition
in the vicinity of $\rho=0.5$, our results on this point are not
conclusive. While the power-law distribution and the long time scales
suggests a nearby critical point, it is not certain the system in
thermodynamic limit indeed reaches this point. 

We thank Alexander Grosberg and Paul Chaikin for interesting and useful
discussions. DL acknowledges support from Israel Science Foundation
grant 1254/12 and US - Israel Binational Science Foundation grant
2008483. DH thanks the US - Israel Binational Science Foundation for
an R. Rahamimoff Travel Grant. We would like to thank the Center for
Soft Matter Research at NYU and the Initiative for Theoretical Science
of the Graduate Center of CUNY for their hospitality while this work
was being conducted.

\bibliographystyle{apsrev}
\bibliography{bib}

\end{document}